# Simulation of Schottky-Barrier Phosphorene Transistors

Runlai Wan, Xi Cao and Jing Guo
Department of Electrical and Computer Engineering
University of Florida, Gainesville, FL, 32611

**Abstract**

Schottky barrier field-effect transistors (SBFETs) based on few and mono layer phosphorene are simulated by the non-equilibrium Green's function formalism. It is shown that scaling down the gate oxide thickness results in pronounced ambipolar I-V characteristics and significant increase of the minimal leakage current. The problem of leakage is especially severe when the gate insulator is thin and the number of layer is large, but can be effectively suppressed by reducing phosphorene to mono or bilayer. Different from two-dimensional graphene and layered dichalcogenide materials, both the ON-current of the phosphorene SBFETs and the metal-semiconductor contact resistance between metal and phosphorene strongly depend on the transport crystalline direction.

Atomically thin two-dimensional (2D) materials, such as graphene and layered dichalcogenide materials, have been intensively studied for potential device applications in recent years. Black phosphorus (BP), which has a direct bandgap of ~0.3 eV[1,2] and a measured mobility in the order of ~10,000 cm$^2$/Vs in bulk,[3] has its corresponding atomically thin 2D structures of monolayer and few layer phosphorene. As the thickness reduces from bulk to monolayer, the bandgap stays as direct but its value increases to about ~1.5 eV.[4] Existence of a direct bandgap is preferable for electronic and optoelectronic device applications, and field-effect transistors (FETs) based on multilayer phosphorene have been experimentally demonstrated. An ON/OFF-current ratio ($I_{ON}/I_{OFF}$) of 5 orders and carrier mobility from several hundred to 1000 cm$^2$/Vs at room temperature (RT) was reported,[5-8] which is likely to be limited by extrinsic impurities. The value, however, already outperforms typical mobility values of transition metal dichalcogenides reported in experiments to date.[9] Theoretical calculations indicate a phonon-limited RT mobility of 10,000 cm$^2$/Vs in monolayer phosphorene for transport along the light effective mass direction.[10] Assessment of performance limits of phosphorene transistors, which assumes ballistic transport and ideal contacts, indicates that highly anisotropic band structure is advantageous and promises considerable ballistic transistor performance improvements over 2D dichalcogenide materials.[11]

In this letter, the device characteristics of Schottky barrier (SB) FETs based on mono and few layer phosphorene are investigated by using device simulation based on the non-equilibrium Green's function (NEGF) formalism.[12] BP FETs demonstrated in experiments to date typically have metal source and drain in direct contact with the channel, which forms Schottky barriers between the source (drain) and the channel. Quantum mechanical tunneling through gate-modulated SBs, which plays an important role in device characteristics of SBFETs, can be described in quantum transport device simulations. It is shown that the ambipolar I-V characteristics, as observed in experiments,[13] can degrade the minimal leakage current ($I_{min}$) considerably as the gate insulator thickness scales down. Reducing the BP thickness to mono or bilayer promises significant advantage over thicker BP materials in terms of reducing the leakage current and improving $I_{ON}/I_{OFF}$ in SBFETs with thin gate insulator. The highly anisotropic band structure of phosphorene plays an important role in tunneling current through the Schottky barriers. The ON-current of phosphorene SBFETs and the contact resistance between metal and phosphorene strongly depend on the transport direction.

Layers of BP are stacked by van der Waals force, and within each layers, phosphorus atoms are formed a puckered honeycomb lattice as shown in Fig. 1(a). For mono and few layer phosphorene, the bands are more dispersive along x direction than along y direction as denoted in Fig. 1(b).[14] The in-plane dispersion of BP with the number of layers (*N*) varying from one to five around the Γ point can be described by a low energy k.p Hamiltonian,[15]

$$H = \begin{pmatrix} E_c + \eta_c k_x^2 + v_c k_y^2 & \gamma k_x + \beta k_y^2 \\ \gamma k_x + \beta k_y^2 & E_v - \eta_v k_x^2 - v_v k_y^2 \end{pmatrix}, \quad (1)$$

where $E_c$ and $E_v$ are the energies of the conduction and valence band edges, $\eta_{c,v}$ and $v_{c,v}$ are related to the effective masses, while $\gamma$ and $\beta$ describe the effective couplings between the conduction and valence bands. These parameters as listed in Table 1 are fitted such that they yield the bandgap and effective mass values obtained by *ab initio* simulation for mono and few layer phosphorene.[10] $E_c/E_v$ is chosen as $\pm E_g/2$. A double-gated SBFET with intrinsic mono and few layer phosphorene as the channel material, which is depicted in Fig. 1(c), is simulated at RT with a channel length of $L_{ch}$ = 30 nm. Ballistic transport is assumed for simplicity, because of the interest on SB-limited transport and the short channel length. The SB height (SBH) for electron $\phi_B$ used in the simulation is 0.2 eV. The gate oxide thickness $t_{ox}$ = 3 nm with a dielectric constant $\kappa$ = 25. The device parameters are the nominal ones and will be varied for different simulations. To model quantum transport along an arbitrary in-plane direction $x'$ between x and y directions, specified by the angle $\theta$ in Fig. 1(b), a unitary transformation[16] is performed to rotate the coordinates, form $k = (k_x k_y)^T$ to $k' = (k_\parallel k_\perp)^T$, $k' = Uk$

$$, where \quad U = \begin{pmatrix} \cos\theta & \sin\theta \\ \sin\theta & -\cos\theta \end{pmatrix}. \quad (2)$$

And a new Hamiltonian *H'* can be obtained by substituting $k_x, k_y$ in Eq. (1) with $k_\parallel$ and $k_\perp$.

The NEGF formalism is used to perform quantum transport calculations with a phenomenological contact self-energy $\Sigma_{S,D} = -it_0$ to mimic the continuous carrier injection from metal to the channel. The nominal value of $t_0$ is 2 eV and variation of the value does not change qualitative conclusions. The mode space approach[17] is applied in the transverse direction and the charge and current are computed by summing over transverse modes. The transverse wave vector ($k_\perp$) in the Hamiltonian is specified as a number determined by the mode index and the wave vector ($k_\parallel$) along the transport direction is treated as an operator and discretized. For the *i*th mode, the net charge density is calculated by integrating the local density-of-states (LDOS) over energy.

To treat self-consistent electrostatics, the NEGF transport equation is self-consistently solved with a 2D Poisson equation[18] in the cross section of the transistor as shown in Fig. 1(c). The boundary conditions in the Poisson equation for source (drain) contact and gate contacts are set to be the Dirichlet boundary condition. When the self-consistency between the quantum transport equation and Poisson equation is achieved, the source-drain current per width *W* is computed by

$$I = \frac{2e}{h}\frac{1}{W}\sum_i \int dE \cdot T_i(E)[f(E - E_{FS}) - f(E - E_{FD})], \qquad (3)$$

where $T_i(E)$ is the source-drain transmission for the *i*th mode calculated by the NEGF formalism.[12]

The BP FETs demonstrated in experiments typically have a channel of multilayer phosphorene. We first simulate the I-V characteristics of a multilayer BP SBFET with 5 layer (5L) phosphorene channel. The simulated $I_D$ vs $V_G$ characteristics show an ambipolar behavior, which is due to electron conduction in one branch and hole conduction in the other, qualitatively agrees with recent experiments on BP FETs.[5,13] The ambipolar characteristics have been studied before in the context of carbon nanotube SBFETs, and is known to degrade the leakage current.[19] On the other hand, to improve the gate electrostatic control, a thin gate oxide thickness is preferred. To study the effect of gate insulator scaling, the $I_D$ vs $V_G$ characteristics at different $t_{ox}$ from 15 nm to 3nm are shown in Fig. 2(a). Scaling down the gate insulator thickness leads to orders of magnitude increase of $I_{min}$ and impacts the capability to turn off the transistor. As $t_{ox}$ decreases from 15 nm to 3 nm, $I_{min}$ increases from 0.18 µA/µm to 12.7 µA/µm. With the oxide thickness scaling down, a better gate control makes the SB thinner, which facilitates the tunneling of carriers. This is further illustrated in Fig. 2(b), which plots the band profile and corresponding transmission when BP SBFETs is operating at ambipolar point for $t_{ox}$ = 3 nm (blue dashed line) and $t_{ox}$ = 15 nm (red solid line) cases. It shows that much thicker SBs effectively block the tunneling of carriers when $t_{ox}$ = 15 nm. Figure 2(a) also shows that thin oxide is preferable for large ON-current. The 5L phosphorene SBFETs with thin gate-oxide thickness, however, suffer severe leaking problem because of strong ambipolar conduction lead by small bandgap ($E_g$ = 0.52 eV).

The leakage problem, which is severe when the gate insulator is thin and the number of layers is large, can be suppressed with reduced number of phosphorene layers. Although the bandgap of bulk BP is relatively small, the value increases substantially when $N$ is reduced especially for $N < 5$ as shown in Table 1. The effect of $N$ on transistor performance is examined next, as shown in Fig. 2(c). Using smaller $N$ enlarges the bandgap, resulting in a significantly reduced $I_{min}$. More specifically, $I_{min}$ goes down below $10^{-2}$ µA/µm when $N < 3$ and reaches $10^{-9}$ µA/µm for monolayer phosphorene, 10 orders of magnitude improvement from the value of 5L. At the same time, though the ON-current decreases with smaller $N$, the changes are within 3 orders (inset of Fig. 1(b)), which means a substantial increase of $I_{ON}/I_{OFF}$ can be achieved. The significant improvement can be explained by Fig. 2(d), where the band profile and corresponding transmission for monolayer phosphorene SBFET operating at ambipolar point are plotted. Compared to 5L case (solid line in Fig.2(b)), the transmission probabilities do not change much, but electron (hole) density in conduction (valence) band decreases due to larger bandgap of monolayer phosphorene ($E_g = 1.51$ eV), which gives a smaller $I_{min}$.

The band structure of phosphorene is highly anisotropic, and the I-V characteristics along different transport directions are examined next. It has been shown that the ballistic performance limits of BP FET[11] and mobility values[10] along x direction and y direction are drastically different. The device characteristics of SBFETs, however, are determined by quantum tunneling transport through gate modulated SBs, which is different from semiclassical diffusive or ballistic transport. Figure 3(a) shows the $I_D$ vs. $V_G$ characteristics of monolayer phosphorene SBFETs with different transport directions, where $\theta$ is defined before. The largest $I_{ON}$ for both electron and hole conduction branches are obtained in the x direction ($\theta = 0°$) due to the lightest effective mass in x direction ($m^*_{ex} = 0.17m_0$ and $m^*_{hx} = 0.15m_0$). On the other hand, the device with channel along y direction ($\theta = 90°$) provides the smallest $I_{min}$ caused by low tunneling transmission due to the heaviest effective mass ($m^*_{ey} = 1.12m_0$ and $m^*_{hy} = 6.35m_0$). The current transporting in direction between x and y directions decreases continuously with the increasing of $\theta$, which agrees well with the gradually changed effective mass from x direction to y direction calculated by *ab initio* simulation.[20] Because the tunneling probability is exponentially sensitive to the value of effective mass, both the minimal leakage current and the ON-current are sensitive to the channel crystalline direction. For a fair comparison of the ON-current, a common OFF-current of $I_{OFF} = 0.1$ µA/µm and power supply voltage of $V_{DD} = 0.5$ V are specified. The gate voltage that leads to $I_{OFF}$ at the electron conduction branch is identified as $I_D$ ($V_G = V_{OFF}$, $V_D = V_{DD}$) = $I_{OFF}$. The ON-current is defined as the gate voltage increases by $V_{DD}$, i.e. $I_{ON} = I_D$ ($V_G = V_{OFF} + V_{DD}$, $V_D = V_{DD}$). The OFF-state gate voltage, $V_{OFF}$ could be switched to 0 V in transistor threshold voltage design by, for example, gate work function engineering. Then the polar representation of $I_{ON}$ is plotted in Fig. 3(b). It clearly shows that $I_{ON}$ is strongly dependent on transport direction and increases monotonically from the zigzag y direction (25 µA/µm) to the armchair x direction (420 µA/µm).

Figure 4(a) shows the $I_D$ vs. $V_D$ characteristics for monolayer phosphorene SBFETs with different transport direction at ON-state which is defined before. $\phi_B$ is set to 0.2 eV in this simulation. As expected, $I_D$ is strongly dependent on transport direction. Transporting along the direction closer to x direction provides a larger current. The anisotropic band structure makes the contact resistance between the phosphorene and metal highly dependent on the transport direction. For the modeled ballistic SBFETs with Schottky barriers at two ends of the channel, the transistor channel resistance at ON-state $R_{ON}$, which is calculated as the inversed of the slope of $I_D$-$V_D$ curve near $V_D = 0$ V point when the gate voltage is at the ON-state value, is limited by the metal-semiconductor contacts between the source (drain) and the channel, rather than any scattering events in the channel. The value of $R_{ON}$ therefore is determined by the serial effect of two metal-semiconductor contacts at the source and drain ends. The results are shown in Fig. 4(b) (circle). $R_{ON}$ is also highly dependent on the choice of transport direction, it is about 10 times smaller if transporting along x direction (493 Ωµm) than along y direction (4536 Ωµm). To get a smaller $R_{ON}$, lower SBH, $\phi_B = 0.1$ and 0.05 eV are also applied and the corresponding $R_{ON}$ values are shown in Fig. 4(b). The direction dependency is similar to $\phi_B = 0.2$ eV case, however, when $\phi_B = 0.05$ eV, a relatively small $R_{ON}$, 206 Ωµm, can be obtained if the light effective mass x direction is chosen as transport direction.

In summary, mono and few layer phosphorene SBFETs are investigated by self-consistent NEGF simulations. Due to a relatively small bandgap, the problem of leakage due to ambipolar transport can be severe if the gate insulator is scaled down and the number of layer is large, but it can be reduced by several orders of magnitude in mono and bilayer phosphorene channel. The highly anisotropic band structure of mono and few layer phosphorene suggests that proper choice of transport direction is important in optimizing the minimum leakage current, ON-current and contact resistance.

The work was supported by ONR and NSF.

**Figure Captions**

Figure 1. (a) Atomic structure of few layer phosphorene marked with coordinate axes ($x, y, z$). (b) Top view of the atomic structure of monolayer phosphorene marked with coordinate axes ($x, y$). Transport direction $x'$ used in simulation is defined by the angle $\theta$ with respect to $x$ direction. (c) Double-gated BP SBFET with metal source and drain contacts. Both the gate and the channel length are 30 nm. The gate oxide thickness $t_{ox}$ = 3 nm with a dielectric constant $\kappa$ = 25.

Figure 2. (a) $I_D$ vs. $V_G$ of five layer phosphorene SBFETs. The gate oxide thickness $t_{ox}$ varies from 15 nm down to 3 nm. (b) Band profile and corresponding transmission for five layer phosphorene SBFETs operating at ambipolar point for $t_{ox}$ = 3 nm (blue dashed line) and $t_{ox}$ = 15 nm (red solid line) cases. Source Fermi level ($\mu_S$) and drain Fermi level ($\mu_D$) are also marked. (c) $I_D$ vs. $V_G$ of 3 nm oxide phosphorene SBFETs with different number of layers. The inset shows the linear scale of $I_D$-$V_G$ characteristics. (d) Band profile and corresponding transmission for monolayer phosphorene SBFETs operating at ambipolar point with $t_{ox}$ = 3 nm. The transistor channel is along x direction as denoted in Fig. 1(b) and the modeled device structure is shown in Fig. 1(c).

Figure3. (a) $I_D$ vs. $V_G$ characteristic of monolayer phosphorene SBFETs with different channel transport direction. The oxide thickness $t_{ox}$ is 3 nm. (b) Polar representation of $I_{ON}$. The in-plane x and y directions are denoted in Fig. 1(b).

Figure 4 (a) $I_D$ vs. $V_D$ characteristics for monolayer phosphorene SBFETs with different channel transport direction at ON-state. The oxide thickness $t_{ox}$ is 3 nm. (b) Polar representation of the channel resistance $R_{ON}$ for $\phi_B$= 0.2, 0.1 and 0.05 eV. $R_{ON}$ is calculated as the inverse of the slope of $I_D$-$V_D$ curve near $V_D$ = 0 V point, which illustrated as dashed line in Fig. 4a. The inset shows $R_{ON}$ as a function of $\theta$ when $\phi_B$= 0.05 eV.

**Table Captions**

TABLE 1. Parameters used in the Hamiltonian of BP with number of layer varying from one to five, fitted to the bandgap and effective mass calculated by *ab initio* simulation.[10] Subscript 'e' and 'h' denote 'electron' and 'hole', and subscript 'x' and 'y' denote x direction and y direction respectively as shown in Fig. 1(b).

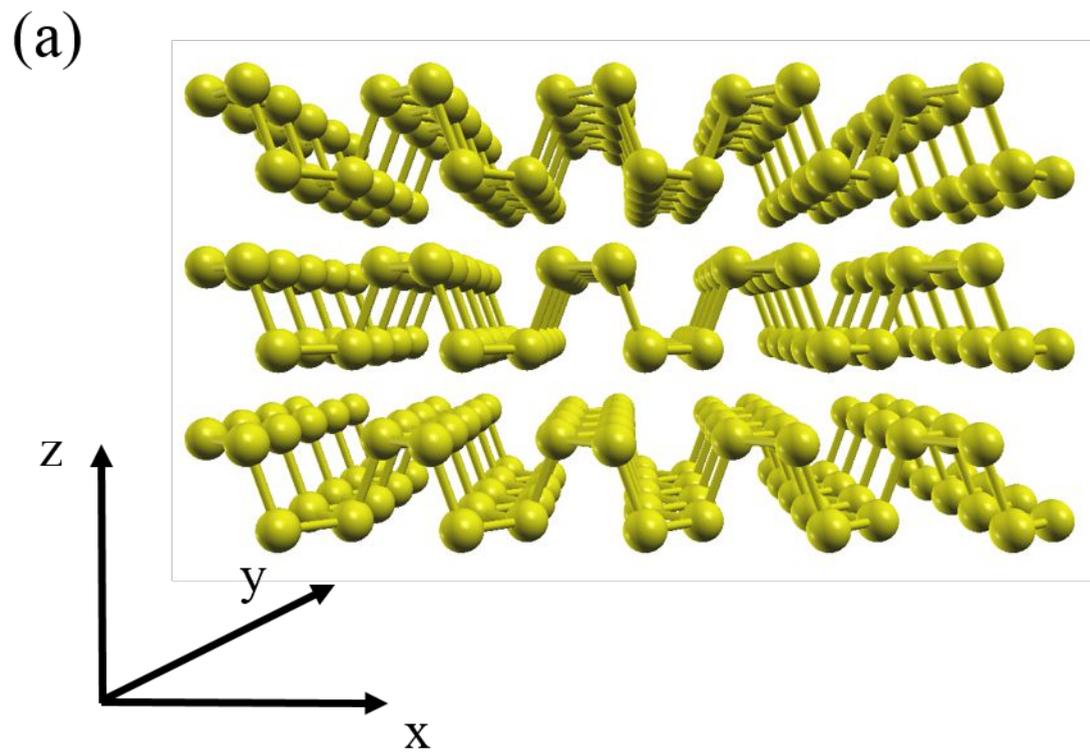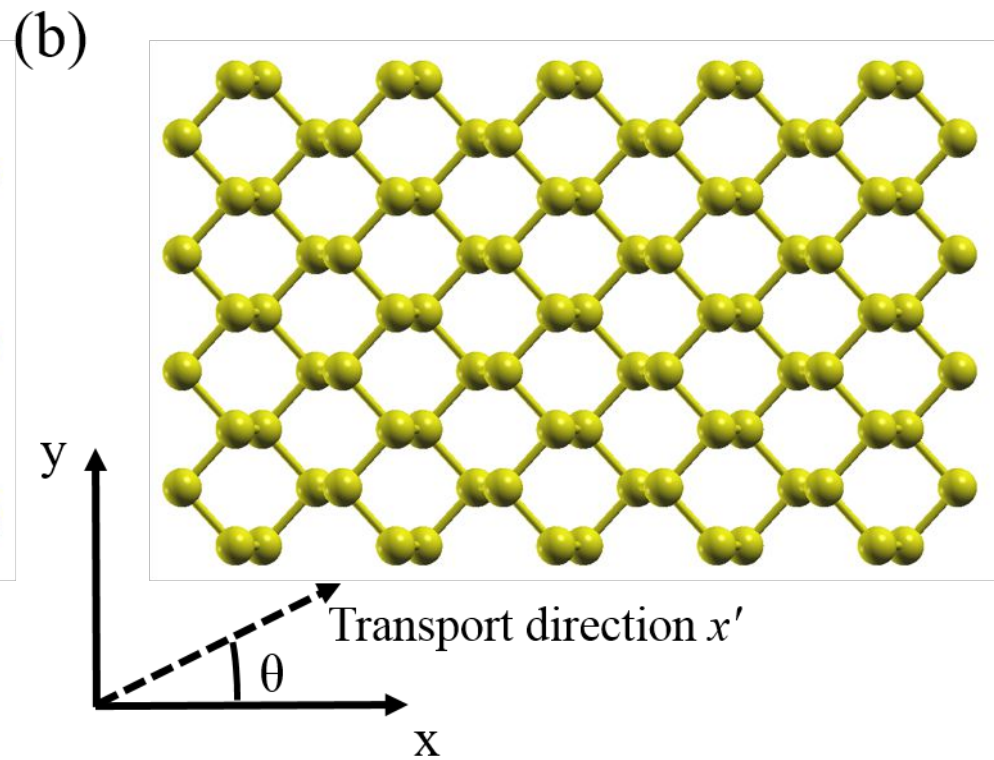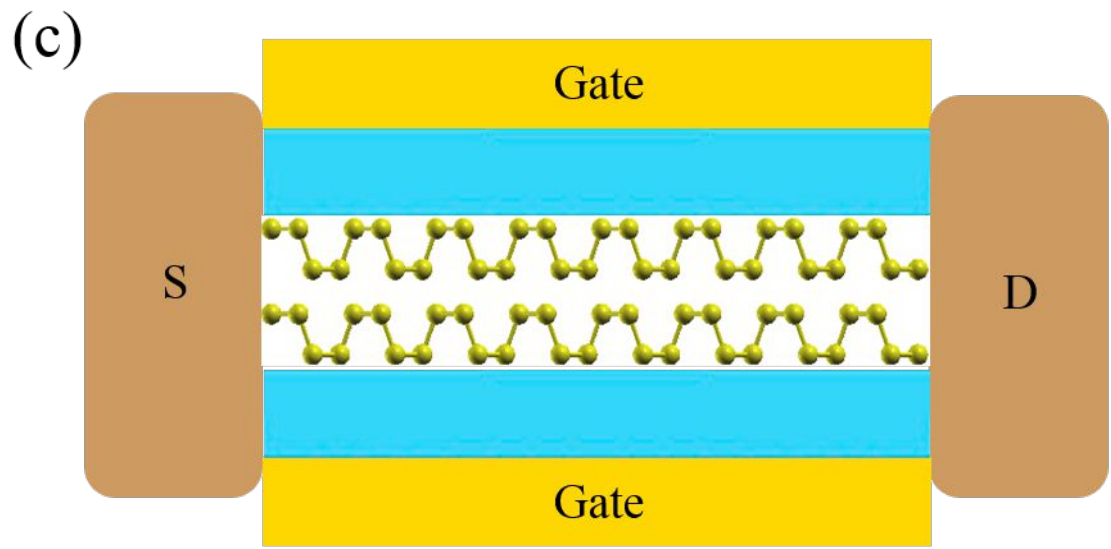

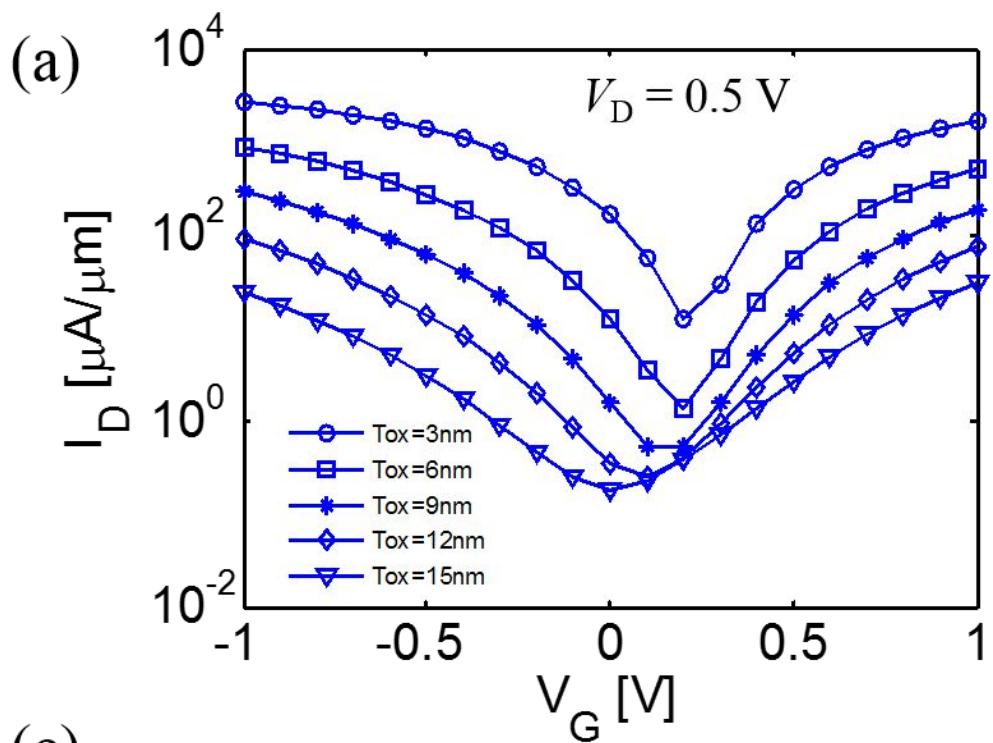
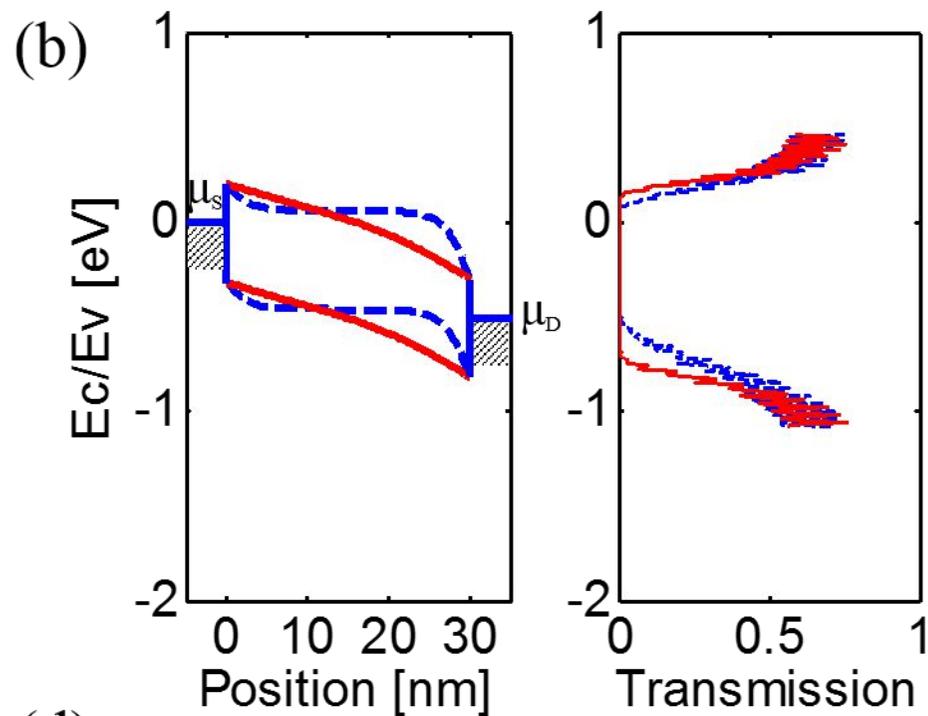
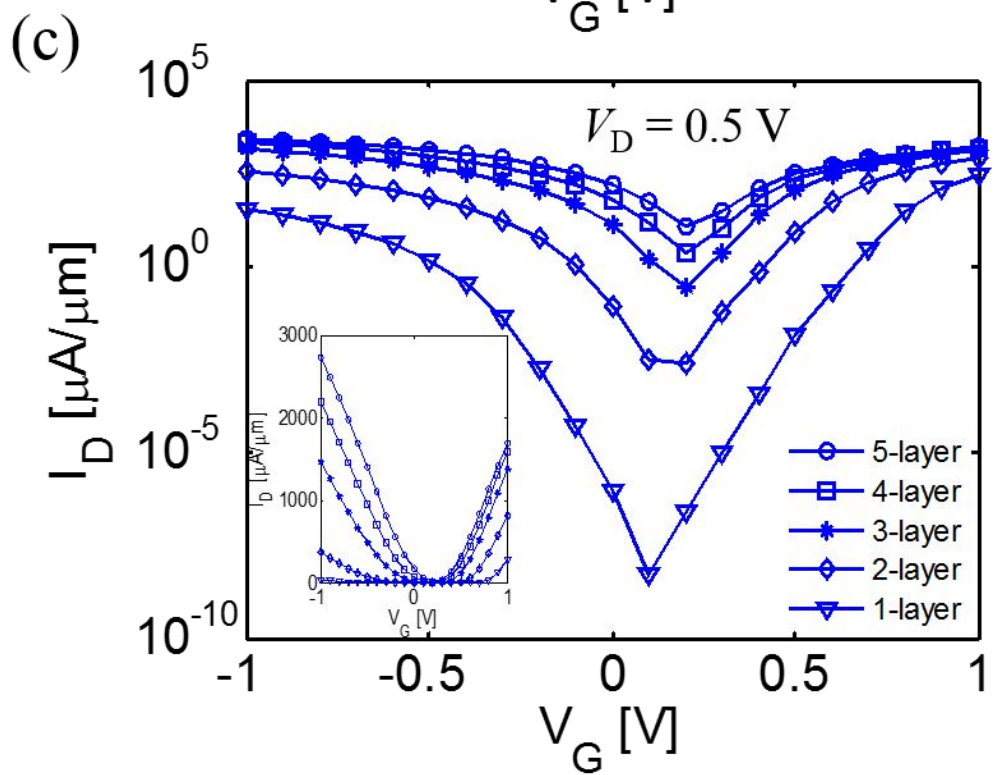
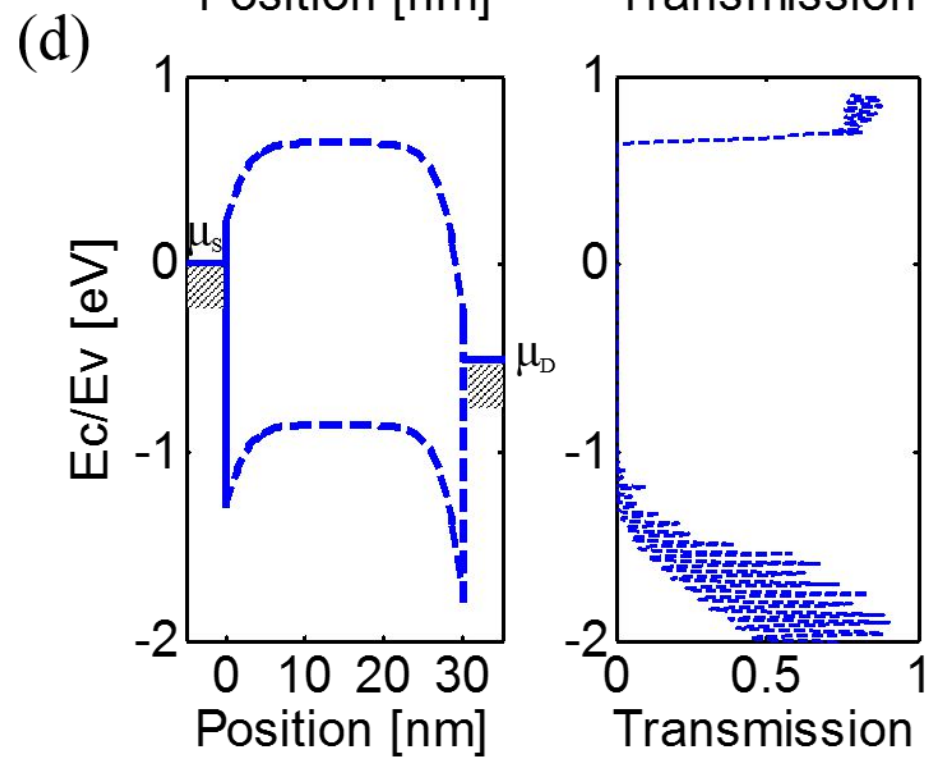

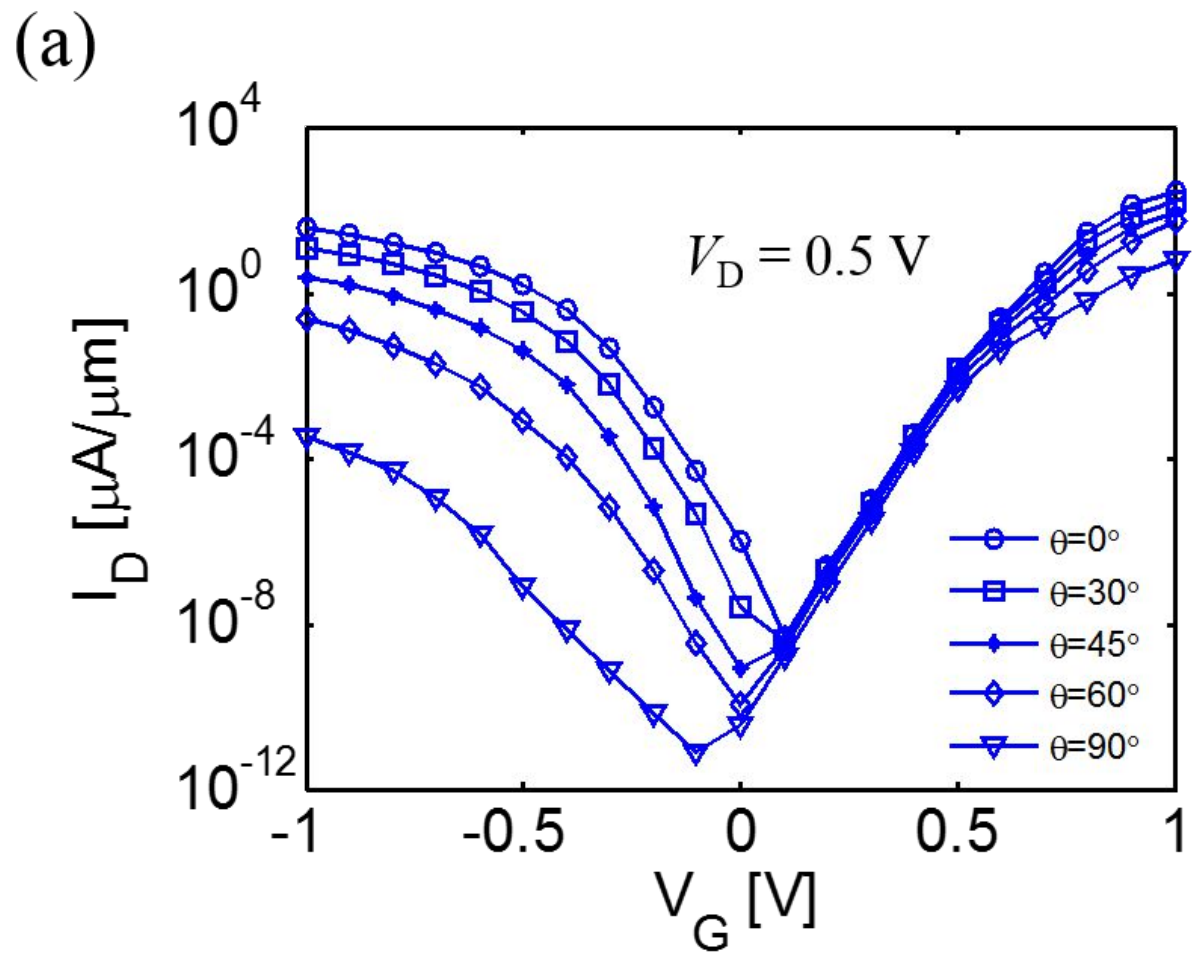 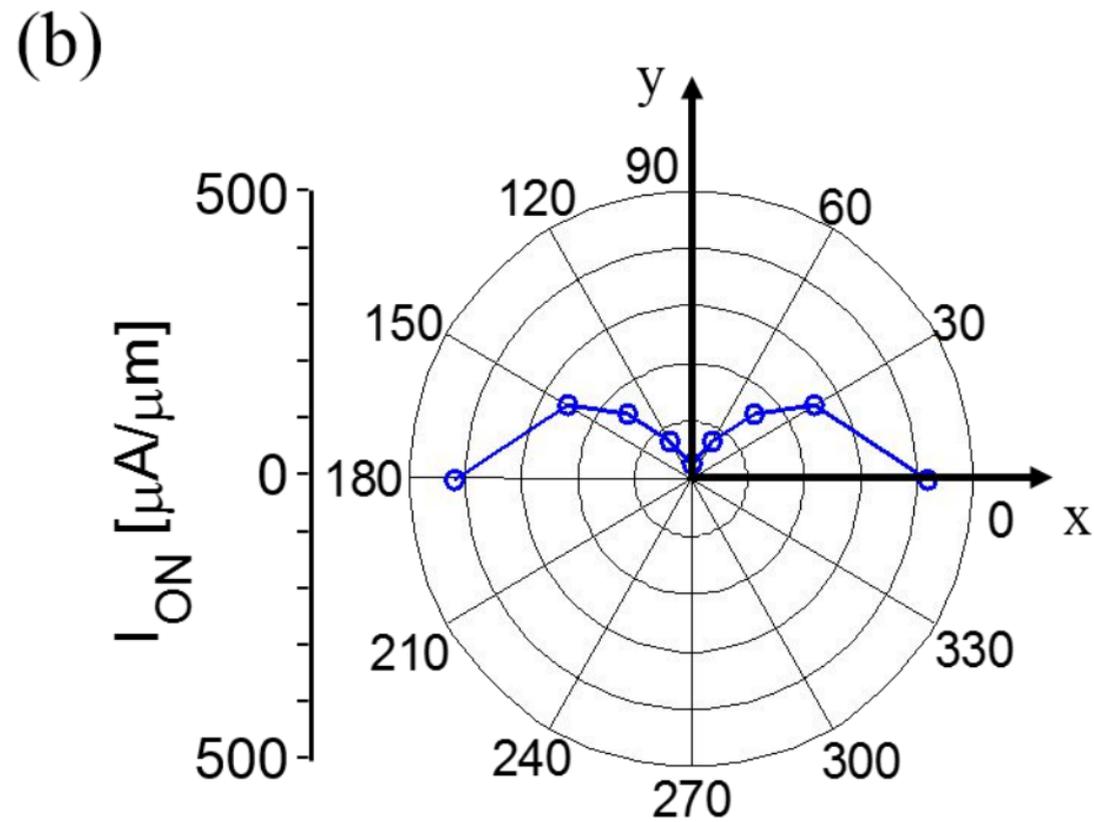

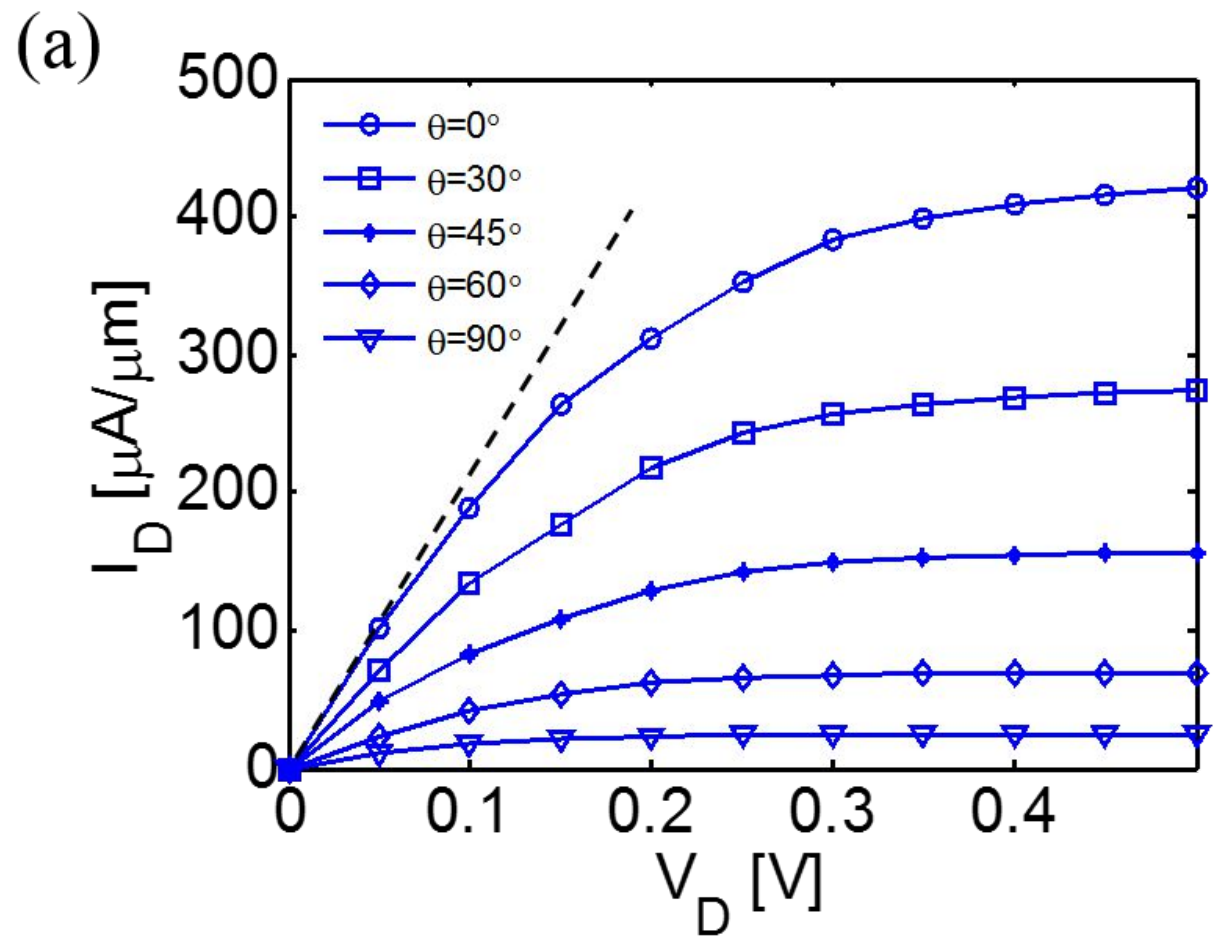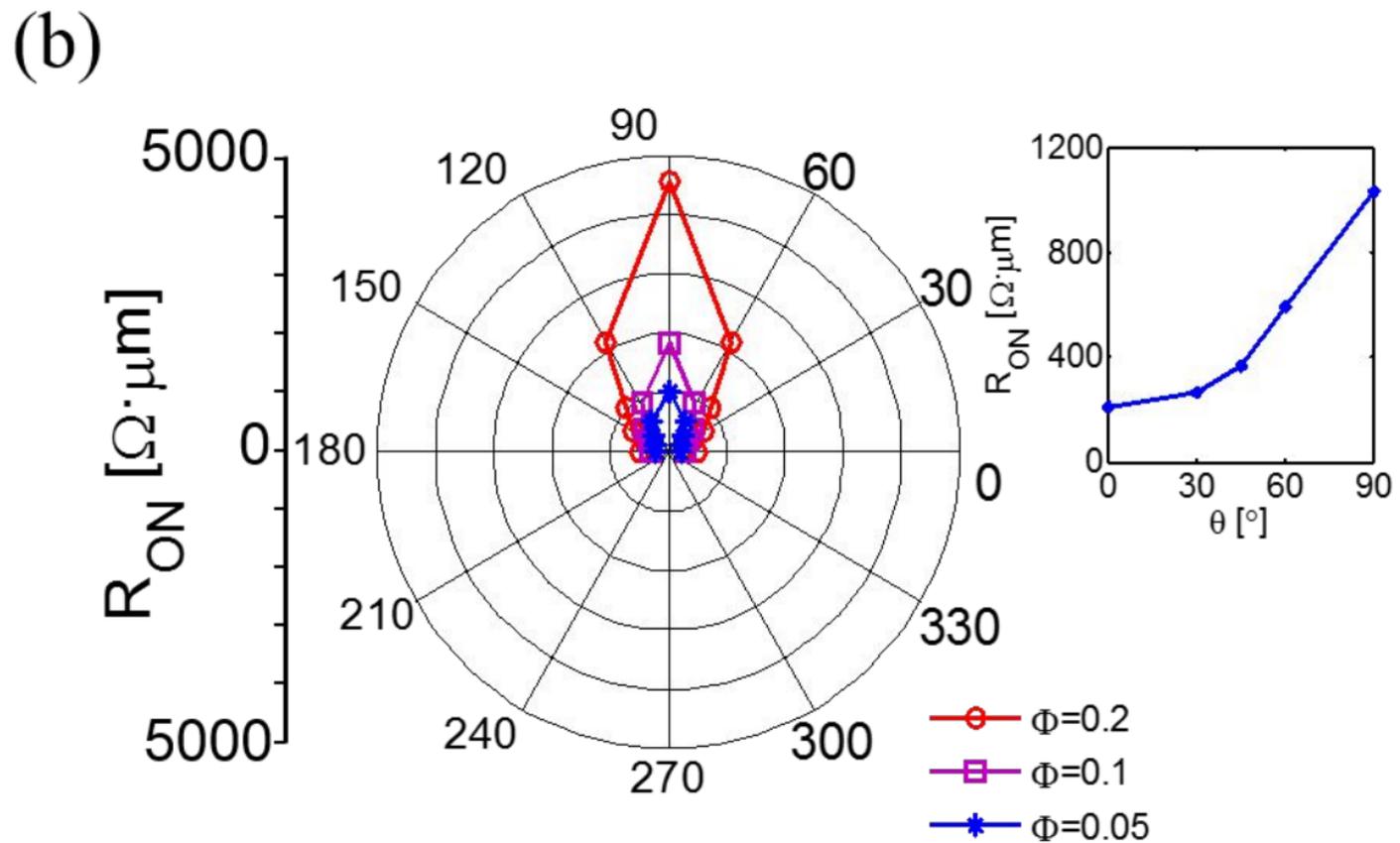

| # of layers | $\eta_c$ (eVÅ$^2$) | $\eta_v$ (eVÅ$^2$) | $v_c$ (eVÅ$^2$) | $v_v$ (eVÅ$^2$) | $\gamma$ (eVÅ) | $\beta$ (eVÅ$^2$) | $E_g$ (eV) | $m_{ex}/m_0$ | $m_{ey}/m_0$ | $m_{hx}/m_0$ | $m_{hy}/m_0$ |
|---|---|---|---|---|---|---|---|---|---|---|---|
| 1 | 3.82 | 3.82 | 3.18 | 0.59 | 5.81 | 4.47 | 1.51 | 0.17 | 1.12 | 0.15 | 6.35 |
| 2 | 1.27 | 5.45 | 3.38 | 2.11 | 4.50 | 4.47 | 1.02 | 0.18 | 1.13 | 0.15 | 1.81 |
| 3 | 2.73 | 4.24 | 3.32 | 3.41 | 3.92 | 4.47 | 0.49 | 0.16 | 1.15 | 0.15 | 1.12 |
| 4 | 2.55 | 5.45 | 3.29 | 3.94 | 3.63 | 4.47 | 0.67 | 0.16 | 1.16 | 0.14 | 0.97 |
| 5 | 3.82 | 5.87 | 3.24 | 4.29 | 3.34 | 4.47 | 0.59 | 0.15 | 1.18 | 0.14 | 0.89 |